# Giant nonreciprocal emission of spin waves in Ta/Py bilayers

Jae Hyun Kwon,[1]* Jungbum Yoon,[1]* Praveen Deorani,[1] Jong Min Lee,[1] Jaivardhan Sinha,[2] Kyung-Jin Lee,[3,4] Masamitsu Hayashi,[2,5] and Hyunsoo Yang[1]†

[1]Department of Electrical and Computer Engineering and NUSNNI, National University of Singapore, 117576, Singapore.

[2]National Institute for Materials Science, Tsukuba 305–0047, Japan.

[3]Department of Materials Science and Engineering, Korea University, Seoul 02841, Korea

[4]KU-KIST Graduate School of Converging Science and Technology, Korea University, Seoul 02841, Korea

[5]Department of Physics, The University of Tokyo, Bunkyo, Tokyo 113-0033, Japan

†Corresponding author. E-mail: eleyang@nus.edu.sg

*These authors contributed equally to this work.

**Spin waves are propagating disturbances in the magnetization of magnetic materials. One of their interesting properties is the nonreciprocity, exhibiting that their amplitude depends on the magnetization direction. Nonreciprocity in spin waves is of great interest in both fundamental science and applications, as it offers an extra knob to control the flow of waves for the technological fields of logics and switch applications. We show a high nonreciprocity in spin waves from Ta/Py bilayer systems with out-of-plane magnetic fields. The nonreciprocity depends on the thickness of Ta underlayer which is found to induce an interfacial anisotropy. The origin of observed high nonreciprocity is twofold; different polarities of the in-plane magnetization due to different angles of canted out-of-plane anisotropy and the spin**

**pumping effect at the Ta/Py interface. Our findings provide an opportunity to engineer highly efficient nonreciprocal spin wave based applications such as nonreciprocal microwave devices, magnonic logic gates, and information transports.**

INTRODUCTION

The unidirectional propagation of signals is the key to modern electronic logic circuits and fundamental applications such as a diode, isolator, gyrator, and circulator (*1*). As the nonreciprocity offers an opportunity to control the wave-devices flowing only to a desirable direction, the nonreciprocity in waves has been extensively studied, and wave-devices with nonlinear propagating properties are introduced in fields of microwave (*2*), acoustics (*3, 4*), and photonics (*5, 6*). A typical nonreciprocity ratio of nonreciprocal microwave devices (*7*) is ~$10^3$, and the nonreciprocity of acoustics (*3*) and photonics (*5*) is reported to be ~$10^4$.

In rapidly developing areas of magnonics, the spin waves (*8-14*) can be the information carrier for magnonic logic circuits (*15-19*). The specific property of the nonreciprocity might be a benefit for the enhancement of logic circuits based on magnetic waveguide. Therefore, the characterization and tuning of the nonreciprocity in magnonics has been intensively investigated (*20-31*), including the relation between the nonreciprocity and spin pumping (*32*) or Dzyaloshinskii-Moriya interaction (*33, 34*). The nonreciprocity, due to its intrinsic characteristic in the magnetostatic surface wave (MSSW) mode (*35*), together with Joule-heat-free transport in GHz-frequencies (*30*) and compatibility with the spin Hall current system (*36, 37*) makes spin wave one of the promising candidates for future data carriers.

However, the suitability of nonreciprocal spin waves for the above applications is limited to a small nonreciprocity parameter, $\kappa$. For example, it was reported that a typical value of $\kappa$ in MSSW is 0.6 ~ 0.9 in a permalloy (Py) strip by using time-resolved spin wave spectroscopy (*23, 26*) (see fig. S1), and 4 ~ 5 by using spatially-resolved Brillouin light scattering spectroscopy (*22*). The value of $\kappa$ in a YIG film, which has a low saturation magnetization, is higher compared to that in Py (*27*). However, a large value of $\kappa$ in a typical 3*d*-ferromagnet such as Py is essential for the realization of spin wave based applications, since Py is compatible with prevailing microfabrication technology.

Here, we report giant nonreciprocal emission of spin waves in a Ta/Py bilayer with out-of-plane bias fields with $\kappa$ values up to 14 and 60 in time and frequency domain, respectively. An optical image of the spin wave device used in this study is shown in Fig. 1A. A pulse inductive technique is used for measurements, in which an electric pulse is applied to the right asymmetric coplanar stripline (ACPS) for the excitation of the spin wave in a ferromagnetic thin film (20 nm Py) on top of various thicknesses ($t_{Ta}$) of Ta from 1.2 to 8.9 nm. The spin wave propagating along the *x*-direction is detected using a sampling oscilloscope at the left ACPS. An out-of-plane bias field ($|H_z| < 490$ mT) is applied to the sample during the measurements. All measurements were performed at room temperature.

The nonreciprocity in MSSW is due to either the interference between the spin waves produced by the *x* and *z*-components of *rf* local excitation fields (antenna-spin wave coupling), or the local concentration of the spin waves at the top and bottom surfaces of the magnetic film. In our case, the amplitude of spin waves across the Py thickness can be assumed to be uniform since $kt_{Py} \ll 1$ (*22, 35*), where $k$ ($2\pi/\lambda \sim 0.63$ rad/μm, $\lambda$ is the wavelength of spin waves) is the wave vector and $t_{Py}$ (20 nm) is the thickness of Py. Therefore, the basic mechanism of nonreciprocity in our system is the antenna-spin wave coupling. In addition, we discuss two reasons for the observed high nonreciprocity arising at the Ta/Py interface. The first reason is the tilted out-of-plane anisotropy induced by the Ta underlayer, and the second one is the spin pumping effect at the Ta/Py interface. The vibrating sample magnetometry (VSM) and anisotropic interface magnetoresistance (AIMR) measurements indeed show the presence of an interfacial anisotropy at the Ta/Py interface. Taking this anisotropy into account, the micromagnetic simulations support the observed experimental results.

**RESULTS**

Figure 1B shows the contour plot of the field dependent measurements in time domain for the device with $t_{Ta}$ = 6.1 nm. A representative spin wave packet measured at $H_z$ = 25.8 mT is shown in Fig. 1C for different propagation directions ($\pm k$). The shape of this wave packet closely resembles that of the MSSW. The fast Fourier transform (FFT) of time domain signals in Fig. 1D shows that the travelling spin wave resonance frequency ($f_R$) increases with increasing $|H_z|$ for the device with $t_{Ta}$ = 6.1 nm. Figure 1D also shows that the FFT intensity of spin wave packets is stronger for positive $H_z$, as compared to that for negative $H_z$, demonstrating nonreciprocal emission of spin waves depending on the polarity of the applied out-of-plane fields. For a comparison, the MSSW (in-plane bias fields), the magnetostatic backward volume (MSBV), the magnetostatic forward volume (MSFV), and the MSSW-like (out-of-plane fields) mode of this study are shown in Fig. 1E.

The amplitude of wave packets for devices with different $t_{Ta}$ is shown as a color plot in Fig. 2A. The maximum value of the spin wave amplitude is observed at $t_{Ta}$ = 6.1 nm for the positive $H_z$, and at $t_{Ta}$ = 3.3 nm for the negative $H_z$. Figure 2B shows that the amplitude at $H_z$ = –25.8 mT decreases as $t_{Ta}$ increases, whereas the amplitude at $H_z$ = + 25.8 mT first increases (up to $t_{Ta}$ = 6.8 nm) and then decreases, as summarized in Fig. 2C. The first regime ($t_{Ta}$ < 6.8 nm) is explained by the titled anisotropy induced by Ta layer, and the second regime ($t_{Ta} \geq$ 6.8 nm) is attributed to the spin pumping effect, as discussed later. The spin wave nonreciprocity ($\kappa$) is defined as the ratio of spin wave amplitudes at positive and negative fields, $\kappa = A(+H_z)/A(-H_z)$. The maximum value of $\kappa$ in time domain is ~14 for the device with $t_{Ta}$ = 8.2 nm as shown in the inset of Fig. 2C and the $\kappa$ reaches up to 60 in frequency domain (the details of $\kappa$ in time and frequency domains are documented in Supplementary Materials Section S2 and S3).

We propose that the effect of Ta underlayer is to induce a canted interfacial magnetic anisotropy, which affects the nonreciprocity factor. To confirm the presence of interfacial anisotropy induced by Ta underlayers, the out-of-plane anisotropic interface magnetoresistance measurements (AIMR) are conducted on all the devices and the results are shown in fig. S4C. Due to AIMR, the in-plane electrical resistance in the perpendicular anisotropy materials depends on the out-of-plane component of the magnetization (*38*). In Fig. 3, $\Delta R_{OP}$ at $+H_z$ and $-H_z$ are plotted versus the Ta underlayer thickness, where $\Delta R_{OP}$ is the maximum change in the resistance of the device when the magnetic field is swept along the $z$-direction. $\Delta R_{OP}$ is a measure of AIMR which originates from the Ta/Py interface, due to anisotropic magnetic scattering of electrons (*38*). $\Delta R_{OP}$ changes its polarity across $t_{Ta}$ ~ 3 nm, which suggests a change in anisotropy direction in the devices. It is consistent with the trend observed in a polarity change of spin wave amplitudes in Fig. 2B.

For further confirmation of the presence of induced anisotropy in our sample, vibrating sample magnetometry (VSM) measurements have been conducted, and a weak perpendicular anisotropy in the $z$-direction is indeed observed. The anisotropy is verified from an inner loop from VSM hysteresis as shown in fig. S5, which is in line with a recent report (*39*). The $z$-component of the magnetization shows a finite remanence value ($M_r$) at $H_z = 0$. The measured $M_r$ with rotating the sample around the $x$-axis shows the minimum at a positive angle for $t_{Ta} = 1$ nm, and at negative angles for $t_{Ta} = 3, 5,$ and $10$ nm, indicating a change in the canted angle of perpendicular anisotropy for different $t_{Ta}$. For the thicker $t_{Ta}$, the uniaxial anisotropy is canted to the $y$-axis with the projected $M_y$ along the $-y$-direction with $H_z > 0$, whereas the $M_y$ is along the $+y$-direction for the thinner $t_{Ta}$ with $H_z > 0$.

In order to better understand the relation between anisotropy and observed nonreciprocity, micromagnetic simulations have been performed (see fig. S6). Simulations are carried out on a bilayer structure with the bottom layer having a perpendicular uniaxial anisotropy. The spin wave amplitude obtained from simulations show a qualitative agreement with the measured data as shown in Fig. 2D. From the investigation of magnetization configurations in simulated results, we find that the spin wave nonreciprocity is correlated with the polarity of $y$-component of the magnetization. Since the magnetization has the $y$-component, the observed spin waves are MSSW-like. The tilting direction of the aforementioned anisotropy depends on the $t_{Ta}$ at fixed $\pm H_z$. When the applied $H_z$ varies, the $y$-component of the magnetization changes its sign, leading to a change in relative amplitude of spin waves at $\pm H_z$ (*21-23, 31*). In addition, for a better understanding of the antenna-spin wave (SW) coupling, we spatially and temporally analyzed our micromagnetic simulation results with the relation between the excitation magnetic field from the antenna and the local magnetization at the bottom of the antenna (see fig. S7).

Even though a change in the anisotropy direction depending on $t_{Ta}$ can explain the trend of the nonreciprocity, a high nonreciprocal factor especially at thicker Ta ($t_{Ta} \geq 6.8$ nm) cannot be fully explained using the above model. This discrepancy between the results from the micromagnetic simulations and the experimental observation suggests that a mechanism other than the interfacial anisotropy is also in operation, which becomes more dominant at thicker Ta layers. To further understand the origin of high nonreciprocity, the loss mechanisms for spin wave energy should be considered. Apart from the usual energy loss mechanisms in a ferromagnet, which are quantified by the Gilbert damping $\alpha$, a heavy metal Ta underlayer can also absorb the energy from magnetization dynamics via spin pumping (*32*). Spin pumping is the process in which the magnetization precession in a ferromagnet (Py) generates spin current density $\vec{J}_S \propto \vec{m} \times \left( \partial \vec{m} / \partial t \right)$ into the adjacent paramagnetic layer (Ta) (*40-42*). Due to high spin orbit coupling in Ta, this spin current is converted to a transverse charge current $\vec{J}_C \propto \vec{J}_S \times \vec{m}$ via inverse spin Hall effect (ISHE), which can be measured as a transverse emf across the device.

We have performed spin pumping induced ISHE measurements to investigate energy loss from spin waves in Py to Ta. A schematic of the spin pumping measurement is shown in Fig. 4A. In the geometry of our experiment, a spin pumping induced ISHE signal ($V_{sp}$) can be observed only if there exists a non-vanishing *y*-component of the magnetization at the interface, which accounts for the observation of MSSW-like wave packets in Fig. 2B. The ISHE signals measured at an *rf* frequency of 1.3 GHz corresponding to the spin wave resonance frequency at $|H_z|$ = 25.8 mT are shown in Fig. 4B. The amplitude of spin pumping signals exhibits an abrupt jump at $t_{Ta}$ = 6.8 nm as shown in Fig. 4C, which can be related to the spin diffusion length in Ta (~ 7 nm) (*43*). The amplitude of $V_{sp}$ is a measure of loss of total spin wave energy which is absorbed into the Ta layer (*44*). Thus, above $t_{Ta}$ ~ 6.8 nm, the spin wave amplitude should decrease because of the strong spin pumping effect. In fact in Fig. 2C, this decrease in the spin wave amplitude is clearly observable for both positive and negative applied fields for $t_{Ta} \geq 6.8$ nm.

If the spin-pumping changes damping equally for the +$H_z$ and −$H_z$ spin waves, both wave amplitudes will be reduced proportionately, leading the non-reciprocity to be unchanged. Therefore, the spin pumping must unequally affect the spin waves for two field directions (±$H_z$), in order to explain the observed giant nonreciprocity, especially at thicker Ta ($t_{Ta} \geq 6.8$ nm). We propose a plausible mechanism of the spin wave amplitude dependent damping, in which the spin-pumping-induced Gilbert damping of large-amplitude spin waves at +$H_z$ is smaller than that of small-amplitude spin waves at −$H_z$ (see Supplementary Materials Section S8). Consequently, a large-amplitude (small-amplitude) spin wave experiences less (more) attenuation when propagating, resulting in an increased nonreciprocity at thicker Ta.

In order to confirm the spin pumping effect, we have performed the ferromagnetic resonance (FMR) measurements with the same devices of the spin pumping measurements. Taking the spin pumping effect into account, the presence of Ta layer enhances the damping parameter of Py above the intrinsic value (*45, 46*),

$$\alpha_{\text{eff}} = \alpha_0 + \Delta\alpha = \alpha_0 + \frac{g\mu_B}{4\pi M_s t_{FM}} g_{\text{eff}}^{\uparrow\downarrow},$$

where $\alpha_0$ is the intrinsic damping, $\Delta\alpha$ is an additional damping because of the spin pumping effect, $g$ (2.2) is the g-factor, $\mu_B$ is the Bohr magnetron, $t_{FM}$ (20 nm) is the thickness of the Py film, $M_s$ is the saturation magnetization, and $g_{\text{eff}}^{\uparrow\downarrow}$ is the effective spin-mixing conductance. We have extracted the additional damping constant ($\Delta\alpha$) from the FMR spectra as shown in Fig. 4C. The plotted damping enhancement as a function of the Ta thickness shows good agreement with spin pumping signals.

**DISCUSSION**

From the first principles calculation, 5*d* orbital of Ta which has a large spin-orbit coupling is strongly hybridized with 3*d* orbital of Fe, giving rise to strong perpendicular magnetic anisotropy (PMA) at the interface (*47*). The anisotropy properties such as an axis or strength are sensitive to intermixing, roughness, oxidation, and hybridization at the interface which are affected by different $t_{Ta}$. The origin of the canted perpendicular anisotropy to the *y*-axis could be a change in those parameters depending on $t_{Ta}$. The canted anisotropy only occurs at the Ta/Py interface and the anisotropy direction becomes closer to the *x*-axis away from the interface because of the shape anisotropy. It has been reported earlier that a Ru seed layer on top of NiFe can induce a surface/interface anisotropy (*48*), and Ta or Ru underlayer with CoFeB can result in a perpendicular anisotropy (*49*). In our experiment, the nonreciprocity is found to change with different $t_{Ta}$, while the thickness of Ru is kept constant at 3 nm. Therefore, we conclude that the Ta underlayer induces a progressive change in the spin wave properties.

Finally, we have investigated the possible role of Dzyaloshinskii-Moriya interaction (DMI) in our devices. DMI is observed in a broken inversion symmetry (*50, 51*) and at the interface or surface of magnetic multilayers (*52, 53*), and could be expected to have an opposite effect in the inverted sample structure. Therefore, we perform spin wave measurements in devices with the Ta layer on top of the Py layer (see fig. S9). It turns out that the nonreciprocal ratio in the inverted devices show a similar trend as that of Ta underlayer structures. We conclude that the structural sequence between the Ta and Py layer does not affect the nonreciprocal emission of spin waves in our case.

It must be noted that our observations are different from the earlier reports on spin wave nonreciprocity studied only in the MSSW mode (*22, 23, 26, 28*). In order to compare our results with the MSSW mode, we have also measured the nonreciprocity value with sweeping in-plane magnetic field, $H_y$. It is found that a small value (~ 0.7) of nonreciprocity is observed regardless of the Ta thickness similar to previous reports (*54*) (see fig. S1).

The observed very large nonreciprocity of spin waves in the Ta/Py bilayer can be attributed to the combined effects such as the polarity change in the in-plane magnetization due to the canted out-of-plane anisotropy and the spin pumping process at the Ta/Py interface. Utilizing an intimate coupling between the interfacial spin configuration and spin pumping effect, even larger nonreciprocity can be achieved. In addition, non-collinear spin configuration due to DMI may enhance the nonreciprocity ratio (*33*). Our intriguing findings will be useful for future spin wave-based applications such as spin wave information transports and logic devices due to the high nonreciprocal ratio.

## MATERIALS AND METHODS

### Sample preparation

Si/SiO$_2$ substrates/Ta ($t_{Ta}$ nm)/Ni$_{81}$Fe$_{19}$ (20 nm)/Ru (3 nm) was patterned in a rectangular shape (120 × 340 μm$^2$). Magnetic fields were not applied during the sputtering. The wafer was being rotated during sputtering to avoid a field deposition under any stray field effects from the magnetron sputtering gun or the chamber. The SiO$_2$ (35 nm) insulating layer was deposited to isolate an ACPS from the Py layer. The ACPS consists of Ta (5 nm)/Au (100 nm). The signal line width is 10 μm. The gap (edge to edge) between the signal line and the ground conductor is 10 μm, and the distance between the signal lines of two ACPSs is 10 μm as shown in Fig. 1A.

### Measurement set-up

We have used pulsed inductive microwave magnetometry (PIMM) to generate and measure the spin wave signals. Spin waves were generated by applying a 1.8 V electric pulses to the excitation ACPS to produce local *rf* magnetic fields around the signal line. The rise and fall times of the electric pulse were ~70 and ~80 ps, respectively. The pulse width was ~100 ps and the pulse cycle frequency was 100 MHz. The excited spin waves propagate through the Py strip and were inductively detected by the detection ACPS (*8, 55*). The detected spin wave signals were then amplified using a low noise amplifier (29 dB gain) and were subsequently passed into a sampling oscilloscope. The waveforms were averaged 10,000 times by the oscilloscope to improve the signal-to-noise ratio, followed by subtraction of reference signal to obtain the pure wave packets. To obtain the additional damping constant, vector network analyzer based ferromagnetic resonance measurements were performed by a frequency-swept technique with the same devices used in the spin pumping measurements.

**Simulations**

We used object oriented micromagnetic framework (OOMMF) developed by M. J. Donahue and D. G. Porter (see http://math.nist.gov/oommf/) to simulate spin waves by solving the Landau-Lifshitz-Gilbert equation. The magnetic film structure used in our simulations consists of two Py layers (top and bottom layer) with different anisotropy axes. The anisotropy axis was titled 0.4° away from $z$-axis towards $-y$ direction for thick $t_{Ta}$ and towards $+y$ direction for thin $t_{Ta}$. The value of the uniaxial anisotropy constant ($K_u$), saturation magnetization ($M_s$) and exchange stiffness constant ($A$) were $7.0 \times 10^5$ J/m$^3$, $8.0 \times 10^5$ A/m, and $1.3 \times 10^{-11}$ J/m, respectively. The Gilbert damping constant ($\alpha$) and the gyromagnetic ratio ($\gamma$) in our simulations were set to 0.01 and $2.32 \times 10^5$ m/(A·s), respectively. The dimensions of top and bottom layers were $8000 \times 200 \times 16$ nm$^3$ and $8000 \times 200 \times 4$ nm$^3$, respectively and the simulation cell size was $4 \times 200 \times 4$ nm$^3$. To excite spin wave dynamics, a magnetic field pulse of amplitude 3 mT was applied to a volume of $200 \times 200 \times 20$ nm$^3$ at a distance of 1 $\mu$m from one end of Py strip, and the propagating spin waves were detected by the summing the dynamic magnetization over a volume of $800 \times 200 \times 20$ nm$^3$ at 5 $\mu$m away from the excitation source. The out-of-plane bias field was varied from –500 to +500 mT during our simulations.

**SUPPLEMENTARY MATERIALS**

Section S1. Nonreciprocal ratios of surface spin wave mode with in-plane fields.

Section S2. Nonreciprocity ratios from spin wave amplitudes in time domain.

Section S3. Nonreciprocity from spin wave intensity in frequency domain.

Section S4. AMR measurements.

Section S5. Vibrating sample magnetometry (VSM) measurements.

Section S6. Simulations: out-of-plane hysteresis and nonreciprocity.

Section S7. Antenna-spin wave coupling.

Section S8. Spin wave-amplitude-dependent damping.

Section S9. Ta on top of Py layer.

Section S10. Dependence of nonreciprocity on different directions of field sweep.

Section S11. Dependence of nonreciprocity on the sign of the wave vector.

Section S12. Effect of antenna-to-antenna distance on the nonreciprocity.

Fig. S1. Nonreciprocal spin wave emission in MSSW mode.

Fig. S2. Nonreciprocity ratios in time domain.

Fig. S3. Fast Fourier transform (FFT) of spin wave signals.

Fig. S4. AMR measurements.

Fig. S5. Vibrating sample magnetometry measurements.

Fig. S6. Out-of-plane hysteresis and nonreciprocity from simulations.

Fig. S7. Spatial and temporal magnetization excited by the antenna.

Fig. S8. Ta thickness dependence of spin wave resonance frequencies.

Fig. S9. Nonreciprocity in devices with Ta overlayer.

Fig. S10. Nonreciprocity from different directions of field sweep.

Fig. S11. Nonreciprocity from opposite signs of the wave vector.

Fig. S12. Spin wave packets in devices with a longer antenna-to-antenna distance (20 µm).

References (*56-58*)

# REFERENCES AND NOTES

1. D. A. Hodges, H. G. Jackson, *Analysis and Design of Digital Integrated Circuits* (McGraw Hill. New York. ed. 2, 1988).
2. B. K. Kuanr, V. Veerakumar, R. Marson, S. R. Mishra, R. E. Camley, Z. Celinski, Nonreciprocal microwave devices based on magnetic nanowires. *Appl. Phys. Lett.* **94**, 202505 (2009).
3. B. Liang, X. S. Guo, J. Tu, D. Zhang, J. C. Cheng, An acoustic rectifier. *Nature Mater.* **9**, 989-992 (2010).
4. R. Fleury, D. L. Sounas, C. F. Sieck, M. R. Haberman, A. Alù, Sound isolation and giant linear nonreciprocity in a compact acoustic circulator. *Science* **343**, 516-519 (2014).
5. S. J. Ben Yoo, Silicon photonics: A chip-scale one-way valve for light. *Nature Photon.* **3**, 77-79 (2009).
6. L. Fan, J. Wang, L. T. Varghese, H. Shen, B. Niu, Y. Xuan, A. M. Weiner, M. Qi, An all-silicon passive optical diode. *Science* **335**, 447-450 (2012).
7. D. Annapurna, K. D. Sisir, *Microwave engineering* (Tata McGraw Hill. India. 2000).
8. M. Covington, T. M. Crawford, G. J. Parker, Time-resolved measurement of propagating spin waves in ferromagnetic thin films. *Phys. Rev. Lett.* **89**, 237202 (2002).
9. M. L. Schneider, Th. Gerrits, A. B. Kos, T. J. Silva, Gyromagnetic damping and the role of spin-wave generation in pulsed inductive microwave magnetometry. *Appl. Phys. Lett.* **87**, 072509 (2005).
10. K. Perzlmaier, G. Woltersdorf, C. H. Back, Observation of the propagation and interference of spin waves in ferromagnetic thin films. *Phys. Rev. B* **77**, 054425 (2008).
11. G. A. Melkov, Yu. V. Koblyanskiy, R. A. Slipets, A. V. Talalaevskij, A. N. Slavin, Nonlinear interactions of spin waves with parametric pumping in permalloy metal films. *Phys. Rev. B* **79**, 134411 (2009).
12. V. Vlaminck, M. Bailleul, Spin-wave transduction at the submicrometer scale: Experiment and modeling. *Phys. Rev. B* **81**, 014425 (2010).
13. M. A. W. Schoen, J. M. Shaw, H. T. Nembach, M. Weiler, T. J. Silva, Radiative damping in waveguide-based ferromagnetic resonance measured via analysis of perpendicular standing spin waves in sputtered permalloy films. *Phys. Rev. B* **92**, 184417 (2015).
14. P. Wessels, A. Vogel, J.-N. Tödt, M. Wieland, G. Meier, M. Drescher, Direct observation of isolated Damon-Eshbach and backward volume spin-wave packets in ferromagnetic microstripes. *Sci. Rep.* **6**, 22117 (2016).
15. M. P. Kostylev, A. A. Serga, T. Schneider, B. Leven, B. Hillebrands, Spin-wave logical gates. *Appl. Phys. Lett.* **87**, 153501 (2005).
16. A. Khitun, D. E. Nikonov, M. Bao, K. Galatsis, K. L. Wang, Feasibility study of logic circuits with a spin wave bus. *Nanotechnology* **18**, 465202 (2007).
17. T. Schneider, A. A. Serga, B. Leven, B. Hillebrands, R. L. Stamps, M. P. Kostylev, Realization of spin-wave logic gates. *Appl. Phys. Lett.* **92**, 022505 (2008).
18. U. H. Hansen, V. E. Demidov, S. O. Demokritov, Dual-function phase shifter for spin-wave logic applications. *Appl. Phys. Lett.* **94**, 252502 (2009).
19. K. Vogt, F. Y. Fradin, J. E. Pearson, T. Sebastian, S. D. Bader, B. Hillebrands, A. Hoffmann, H. Schultheiss, Realization of a spin-wave multiplexer. *Nat. Commun.* **5**, 3727 (2014).
20. P. K. Amiri, B. Rejaei, M. Vroubel, Y. Zhuang, Nonreciprocal spin wave spectroscopy of thin Ni–Fe stripes. *Appl. Phys. Lett.* **91**, 062502 (2007).
21. T. Schneider, A. A. Serga, T. Neumann, B. Hillebrands, M. P. Kostylev, Phase reciprocity of spin-wave excitation by a microstrip antenna. *Phys. Rev. B* **77**, 214411 (2008).

22. V. E. Demidov, M. P. Kostylev, K. Rott, P. Krzysteczko, G. Reiss, S. O. Demokritov, Excitation of microwaveguide modes by a stripe antenna. *Appl. Phys. Lett.* **95**, 112509 (2009).
23. K. Sekiguchi, K. Yamada, S. M. Seo, K. J. Lee, D. Chiba, K. Kobayashi, T. Ono, Nonreciprocal emission of spin-wave packet in FeNi film. *Appl. Phys. Lett.* **97**, 022508 (2010).
24. M. Kostylev, Non-reciprocity of dipole-exchange spin waves in thin ferromagnetic films. *J. Appl. Phys.* **113**, 053907 (2013).
25. R. Verba, V. Tiberkevich, E. Bankowski, T. Meitzler, G. Melkov, A. Slavin, Conditions for the spin wave nonreciprocity in an array of dipolarly coupled magnetic nanopillars. *Appl. Phys. Lett.* **103**, 082407 (2013).
26. M. Jamali, J. H. Kwon, S.-M. Seo, K.-J. Lee, H. Yang, Spin wave nonreciprocity for logic device applications. *Sci. Rep.* **3**, 3160 (2013).
27. K. L. Wong, L. Bi, M. Bao, Q. Wen, J. P. Chatelon, Y.-T. Lin, C. A. Ross, H. Zhang, K. L. Wang, Unidirectional propagation of magnetostatic surface spin waves at a magnetic film surface. *Appl. Phys. Lett.* **105**, 232403 (2014).
28. P. Deorani, J. H. Kwon, H. Yang, Nonreciprocity engineering in magnetostatic spin waves. *Curr. Appl. Phys.* **14**, S129-S135 (2014).
29. K. Di, S. X. Feng, S. N. Piramanayagam, V. L. Zhang, H. S. Lim, S. C. Ng, M. H. Kuok, Enhancement of spin-wave nonreciprocity in magnonic crystals via synthetic antiferromagnetic coupling. *Sci. Rep.* **5**, 10153 (2015).
30. A. V. Chumak, V. I. Vasyuchka, A. A. Serga, B. Hillebrands, Magnon spintronics. *Nature Phys.* **11**, 453-461 (2015).
31. O. Gladii, M. Haidar, Y. Henry, M. Kostylev, M. Bailleul, Frequency nonreciprocity of surface spin wave in permalloy thin films. *Phys. Rev. B* **93**, 054430 (2016).
32. R. Iguchi, K. Ando, Z. Qiu, T. An, E. Saitoh, T. Sato, Spin pumping by nonreciprocal spin waves under local excitation. *Appl. Phys. Lett.* **102**, 022406 (2013).
33. J.-H. Moon, S.-M. Seo, K.-J. Lee, K.-W. Kim, J. Ryu, H.-W. Lee, R. D. McMichael, M. D. Stiles, Spin-wave propagation in the presence of interfacial Dzyaloshinskii-Moriya interaction. *Phys. Rev. B* **88**, 184404 (2013).
34. F. Garcia-Sanchez, P. Borys, A. Vansteenkiste, J.-V. Kim, R. L. Stamps, Nonreciprocal spin-wave channeling along textures driven by the Dzyaloshinskii-Moriya interaction. *Phys. Rev. B* **89**, 224408 (2014).
35. R. W. Damon, J. R. Eshbach, Magnetostatic Modes of a Ferromagnet Slab. *J. Phys. Chem. Solids* **19**, 308-320 (1961).
36. Z. Duan, C. T. Boone, X. Cheng, I. N. Krivorotov, N. Reckers, S. Stienen, M. Farle, J. Lindner, Spin-wave modes in permalloy/platinum wires and tuning of the mode damping by spin Hall current. *Phys. Rev. B* **90**, 024427 (2014).
37. V. E. Demidov, S. Urazhdin, A. B. Rinkevich, G. Reiss, S. O. Demokritov, Spin Hall controlled magnonic microwaveguides. *Appl. Phys. Lett.* **104**, 152402 (2014).
38. A. Kobs, S. Heße, W. Kreuzpaintner, G. Winkler, D. Lott, P. Weinberger, A. Schreyer, H. P. Oepen, Anisotropic interface magnetoresistance in Pt/Co/Pt sandwiches. *Phys. Rev. Lett.* **106**, 217207 (2011).
39. M. S. Gabor, C. Tiusan, T. Petrisor, Jr., T. Petrisor, The influence of the capping layer on the perpendicular magnetic anisotropy in permalloy thin films. *IEEE Trans. Magn.* **50**, 2007404 (2014).
40. Y. Tserkovnyak, A. Brataas, G. E. Bauer, Enhanced gilbert damping in thin ferromagnetic films. *Phys. Rev. Lett.* **88**, 117601 (2002).

41. K. Ando, Y. Kajiwara, S. Takahashi, S. Maekawa, K. Takemoto, M. Takatsu, E. Saitoh, Angular dependence of inverse spin-Hall effect induced by spin pumping investigated in a Ni(81)Fe(19)/Pt thin film. *Phys. Rev. B* **78**, 014413 (2008).
42. E. Saitoh, M. Ueda, H. Miyajima, G. Tatara, Conversion of spin current into charge current at room temperature: Inverse spin-Hall effect. *Appl. Phys. Lett.* **88**, 182509 (2006).
43. P. Deorani, H. Yang, Role of spin mixing conductance in spin pumping: enhancement of spin pumping efficiency in Ta/Cu/Py structures. *Appl. Phys. Lett.* **103**, 232408 (2013).
44. S. S. Mukherjee, P. Deorani, J. H. Kwon, H. Yang, Attenuation characteristics of spin-pumping signal due to traveling spin waves. *Phys. Rev. B* **85**, 094416 (2012).
45. H. Nakayama, K. Ando, K. Harii, T. Yoshino, R. Takahashi, Y. Kajiwara, K. Uchida, Y. Fujikawa, E. Saitoh, Geometry dependence on inverse spin Hall effect induced by spin pumping in Ni81Fe19/Pt films. *Phys. Rev. B* **85**, 144408 (2012).
46. Y. Tserkovnyak, A. Brataas, G. E. W. Bauer, Spin pumping and magnetization dynamics in metallic multilayers. *Phys. Rev. B* **66**, 224403 (2002).
47. D. Odkhuu, S. H. Rhim, N. Park, S. C. Hong, Extremely large perpendicular magnetic anisotropy of an Fe(001) surface capped by 5d transition metal monolayers: A density functional study. *Phys. Rev. B* **88**, 184405 (2013).
48. L. Jin, H. Zhang, X. Tang, F. Bai, Z. Zhong, Effects of ruthenium seed layer on the microstructure and spin dynamics of thin permalloy films. *J. Appl. Phys.* **113**, 053902 (2013).
49. Y. Shiota, F. Bonell, S. Miwa, N. Mizuochi, T. Shinjo, Y. Suzuki, Opposite signs of voltage-induced perpendicular magnetic anisotropy change in CoFeB|MgO junctions with different underlayers. *Appl. Phys. Lett.* **103**, 082410 (2013).
50. A. Tonomura, X. Yu, K. Yanagisawa, T. Matsuda, Y. Onose, N. Kanazawa, H. S. Park, Y. Tokura, Real-space observation of skyrmion lattice in helimagnet MnSi thin samples. *Nano lett.* **12**, 1673-1677 (2012).
51. X. Z. Yu, N. Kanazawa, Y. Onose, K. Kimoto, W. Z. Zhang, S. Ishiwata, Y. Matsui, Y. Tokura, Near room-temperature formation of a skyrmion crystal in thin-films of the helimagnet FeGe. *Nature Mater.* **10**, 106-109 (2011).
52. S. Heinze, K. von Bergmann, M. Menzel, J. Brede, A. Kubetzka, R. Wiesendanger, G. Bihlmayer, S. Blügel, Spontaneous atomic-scale magnetic skyrmion lattice in two dimensions. *Nature Phys.* **7**, 713-718 (2011).
53. M. Bode, M. Heide, K. von Bergmann, P. Ferriani, S. Heinze, G. Bihlmayer, A. Kubetzka, O. Pietzsch, S. Blügel, R. Wiesendanger, Chiral magnetic order at surfaces driven by inversion asymmetry. *Nature* **447**, 190-193 (2007).
54. J. H. Kwon, P. Deorani, J. Yoon, M. Hayashi, H. Yang, Influence of tantalum underlayer on magnetization dynamics in Ni81Fe19 films. *Appl. Phys. Lett.* **107**, 022401 (2015).
55. T. J. Silva, C. S. Lee, T. M. Crawford, C. T. Rogers, Inductive measurement of ultrafast magnetization dynamics in thin-film Permalloy. *J. Appl. Phys.* **85**, 7849-7862 (1999).
56. M. L. Polianski, P. W. Brouwer, Current-induced transverse spin-wave instability in a thin nanomagnet. *Phys. Rev. Lett.* **92**, 026602 (2004).
57. A. Brataas, Y. Tserkovnyak, G. E. W. Bauer, Current-induced macrospin versus spin-wave excitations in spin valves. *Phys. Rev. B* **73**, 014408 (2006).
58. K.-J. Lee, M. D. Stiles, H.-W. Lee, J.-H. Moon, K.-W. Kim, S.-W. Lee, Self-consistent calculation of spin transport and magnetization dynamics. *Phys. Rep.* **531**, 89-113 (2013).

**Funding:** This work is supported by the National Research Foundation, Prime Minister's Office, Singapore under its Competitive Research Programme (CRP Award No. NRFCRP12-2013-01), Grant-in-Aid for Scientific Research (No. 22760015) from MEXT, Japan, and Creative Materials Discovery Program through the National Research Foundation of Korea (NRF-2015M3D1A1070465). **Author contributions:** J.K., M.H., and H.Y. initiated this work. J.K. and J.-M. L. carried out experiments. J.Y. and P.D. did simulations. M.H. fabricated devices. J.S. did VSM measurements. K-J.L. contributed to spin pumping theory. All authors discussed the results. J.K., J.Y., P.D., and H.Y. wrote the manuscript. H.Y. supervised and led the project. **Competing interest**: The authors declare that they have no competing interests. **Data and materials availability:** All data needed to evaluate the conclusions in the paper are present in the paper and/or the Supplementary Materials. Additional data related to this paper may be requested from the authors.

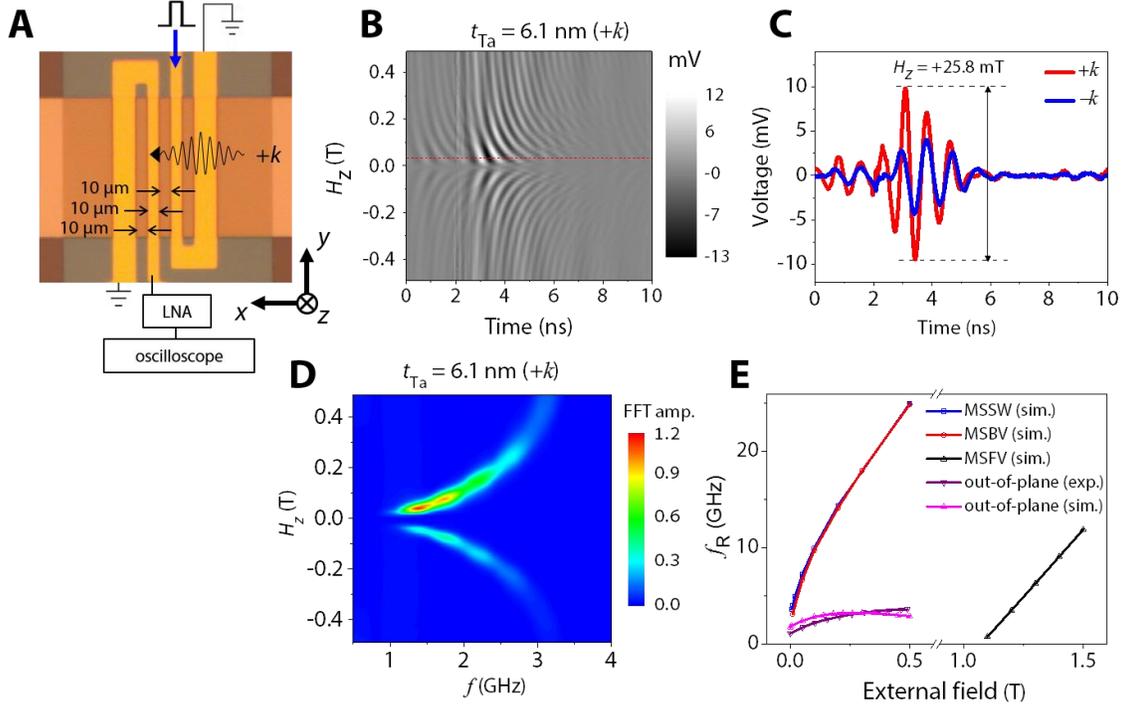

**Fig. 1. Device structure and spin wave signals with out-of-plane bias fields.** (**A**) The device consists of Si/SiO$_2$ substrate/Ta ($t_{Ta}$ nm)/Py (20 nm)/Ru (3 nm)/SiO$_2$ (35 nm). The thickness of Ta underlayer ranges from 1.2 to 8.9 nm. Asymmetric coplanar strips (ACPS) are used to apply local excitation fields and inductively detect magnetization precessions using a sampling oscilloscope after amplification with a low noise amplifier (LNA). (**B**) Contour plot of spin waves in time domain with various $H_z$ for $t_{Ta}$ = 6.1 nm. The fields ($|H_z|$ < 490 mT) are applied perpendicularly to the film plane, which are smaller than the demagnetization field, thus cannot excite MSFV mode. (**C**) Spin wave packet at $H_z$ = 25.8 mT (corresponding to the red dotted line in Fig. 1B). The amplitude is defined as the difference between the maximum and minimum of the wave packet. (**D**) Contour plot of FFT of time domain signals. The spin wave resonance frequency ($f_R$) increases with increasing $|H_z|$. (**E**) The range of $f_R$ corresponding to the external fields differs from that of the conventional spin wave modes, such as MSSW, MSBV, and MSFV. The $f_R$ plots for MSSW, MSBV, MSFV, and out-of-plane (sim.) are obtained from the simulations, in which the material parameters of Py are used. The experimental data of $f_R$ for the out-of-plane (exp.) are obtained from the sample of $t_{Ta}$ = 6.1 nm.

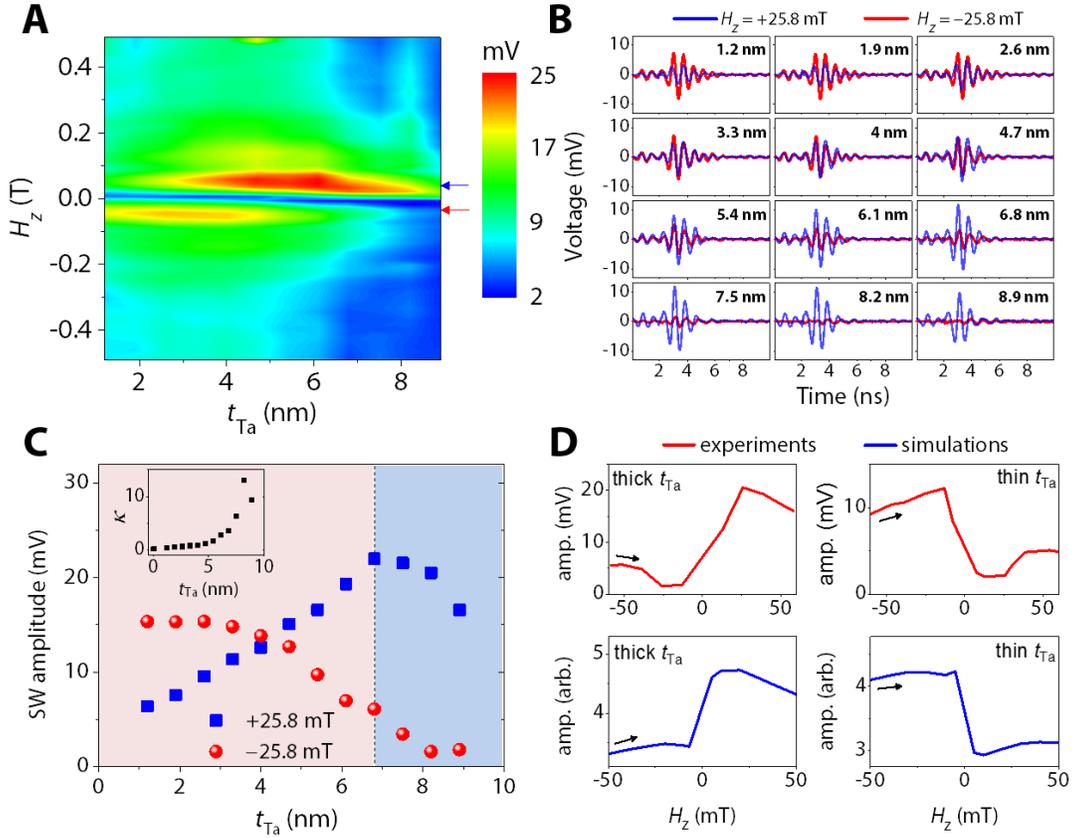

**Fig. 2. Change in spin wave amplitudes with different $t_{Ta}$.** (**A**) Contour plot of spin wave amplitudes as a function of $t_{Ta}$ and $H_z$. The red and blue arrows on the right side indicate $H_z = \pm 25.8$ mT. (**B**) Spin wave packets at $H_z = \pm 25.8$ mT. The amplitude at $-H_z$ is higher than that at $+H_z$ in the device for $1.2 < t_{Ta} < 3.3$ nm. The amplitude at $+H_z$ is higher than that at $-H_z$ for $4.7 < t_{Ta} < 8.9$ nm. (**C**) Amplitudes of the wave packets from Fig. 2B are plotted with different $t_{Ta}$ and the nonreciprocity ($\kappa$) is shown in the inset. $\kappa$ is found to be ~0.2 for $t_{Ta} = 0$ nm. (**D**) Simulation results are shown for comparison with experimental results. $H_z$ sweeps from negative to positive fields. The simulations show a good agreement with measurements. The experimental data are obtained from a thin case ($t_{Ta} = 1.9$ nm) and a thick case ($t_{Ta} = 8.2$ nm).

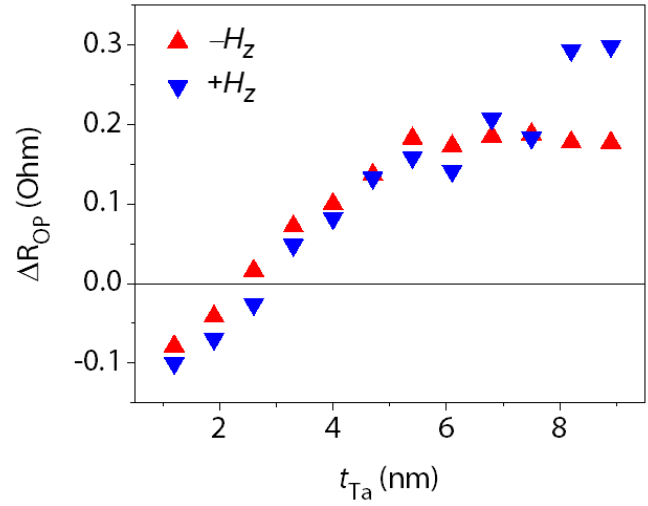

**Fig. 3. Anisotropic interface magnetoresistance.** $\Delta R_{OP}$ is plotted with different $t_{Ta}$. It shows a polarity change at $t_{Ta} \sim 3$ nm and an increase with increasing $t_{Ta}$. It indicates that different $t_{Ta}$ changes the interfacial anisotropy directions.

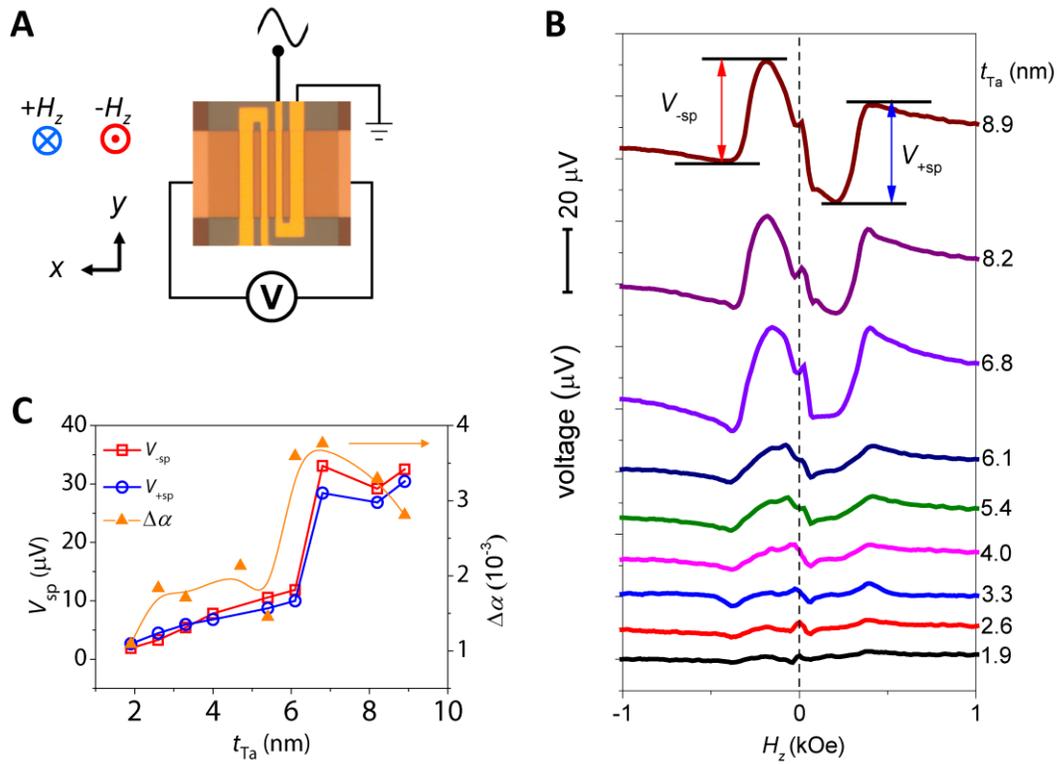

**Fig. 4. Spin pumping measurements by spin waves and ferromagnetic resonance data with out-of-plane $H_z$.** (**A**) The spin pumping voltage is measured at the end of both electrodes. The applied 1.3 GHz microwave continuously excites local magnetization oscillations. (**B**) The amplitude of spin pumping signals increase with thicker $t_{Ta}$. It is expected that the amplitude shows the maximum for $t_{Ta} \sim$ spin diffusion length (7 nm) in Ta. The spin pumping signal with $t_{Ta} = 1.2$ nm is used as a background signal and subtracted from that of all other devices. (**C**) A sudden increase in $V_{sp}$ is observed at $t_{Ta} = 6.8$ nm. It is inferred that the spin pumping effect plays a significant role for $t_{Ta} \geq 6.8$ nm. The plotted damping enhancement ($\Delta\alpha$) as a function of the Ta thickness shows a good agreement with spin pumping signals.